\documentclass[aps,twocolumn]{revtex4}

\usepackage{graphicx}
\usepackage[dvipsnames,usenames]{color}

\begin{document}

\title{
Quantized current blockade and hydrodynamic correlations
in biopolymer translocation through nanopores: evidence from multiscale simulations
}
\author{Massimo Bernaschi$^1$, Simone Melchionna$^2$, Sauro Succi$^1$, 
Maria Fyta$^{3}$, and Efthimios Kaxiras$^3$ }
\affiliation{$^1$ Istituto Applicazioni Calcolo, CNR, 
Viale del Policlinico 137, 00161, Roma, Italy \\
$^2$ INFM-SOFT, Department of Physics, Universit\`a di Roma 
\it{La Sapienza}, P.le A. Moro 2, 00185 Rome, Italy \\
$^3$ Department of Physics and School of Engineering and Applied
Sciences, Harvard University, Cambridge, MA, USA\\
}

\date{\today}

\begin{abstract}
We present a detailed
description of biopolymer translocation through a nanopore in the presence of a
solvent, using an innovative multi-scale methodology which 
treats the biopolymer at the microscopic scale as combined with a
self-consistent mesoscopic description for the solvent fluid dynamics.
We report evidence for quantized current blockade depending on 
the folding configuration and offer detailed
information on the role of hydrodynamic correlations in speeding-up the
translocation process.
\end{abstract}

\maketitle

Biopolymer translocation through nanoscale pores holds
the promise of efficient and improved sensing for many applications 
in biotechnology, 
and possibly ultrafast DNA sequencing~\cite{Kasianowicz_PNAS1996, 
Meller_PNAS2000,ultraseq}.
Recent advances in fabrication of solid-state
nanopores~\cite{Li_Nature2001,Storm_NatMat2003} have spurred 
detailed experimental studies of the translocation
process, with DNA as the prototypical biopolymer of 
interest~\cite{Dekker_Nat2007}. 
Computer simulations that can account for the complexity of the 
biomolecule motion as it undergoes translocation, as well as its 
interaction with the environment
(the nanopore and the solvent), are crucial in elucidating 
current experiments~\cite{Li_NatMat2003,Storm_NanoLett2005} and 
possibly inspiring new ones. 
Here, we study the dynamical, statistical and synergistic features of 
the translocation process
of a biopolymer through a nanopore by a
multiscale method based on molecular dynamics
for the biopolymer motion and mesoscopic lattice Boltzmann dynamics for 
the solvent.
We report evidence for quantized current blockade depending on 
the folding configuration (single- or multi-file translocation) in good 
agreement with
recent experimental observations~\cite{Li_NatMat2003}.
Our simulations show the significance of hydrodynamic correlations
in speeding-up the translocation process.

Nanopores are an essential element of cells and membranes, controlling 
the passage of molecules and regulating many biological processes such as
viral infection by phages and inter-bacterial DNA transduction~\cite{TRANSL}.
The last two decades have witnessed the emergence of artificial 
solid-state nanopores
as potential devices for sensing biomolecules through novel 
means~\cite{Dekker_Nat2007}.
One of the most intriguing possibilities is ultra-fast sequencing of DNA by 
measuring the electronic signal as the biomolecule translocates through a
nanopore decorated with electrodes~\cite{ultraseq}. 
While this goal still remains elusive, a number of
detailed studies on DNA translocation through nanopores have been reported
recently~\cite{Li_NatMat2003,Storm_NanoLett2005}.  These experiments 
typically measure the blockade of the ion current through the 
nanopore during the time it takes the molecule to translocate, which provides
statistical information about the biomolecule motion during the process.

Numerical simulation of the translocation process provides a wealth
of information complementary to experiments, but is hindered 
by the very large number of particles involved in the full process: 
these include 
all the atoms that constitute the biomolecule, the molecules and ions that
constitute the solvent, and the atoms that are part of the solid membrane
in the nanopore region. The spatial and temporal extent of the 
full system on atomic scales 
is far beyond what can be handled by direct computational methods without 
introducing major approximations.
Some universal features of translocation have been analyzed by
means of suitably simplified statistical schemes~\cite{statisTrans},
and non-hydrodynamic coarse-grained or microscopic models
\cite{forrey07,DynamPRL,Nelson} or other mesoscopic approaches
\cite{mesoscSims}.
Many atomic degrees of freedom, and especially
those of the solvent and the membrane wall, are uninteresting from the 
biological  point of view.  The problem naturally calls for a 
multi-scale computational approach that can elucidate the interesting experimental 
measurements while coarse-graining the less important degrees of freedom. 

We have developed a multiscale method for treating the dynamics of biopolymer 
translocation~\cite{ourLBM} and performed an extensive set of numerical simulations, 
combining constrained molecular dynamics (MD) for the polymer motion 
with a Lattice-Boltzmann (LB) treatment of the solvent hydrodynamics \cite{LBE}.
The biopolymer transits through a nanopore under the effect of a localized
electric field applied across the pore, 
mimicking the experimental setup \cite{Storm_NanoLett2005}.
The simulations provide direct computational evidence 
of quantized current blockade and confirm the experimentally surmised  
multiple-file translocation:
the molecule passes through the pore in a multi-stranded fold configuration
when the pore is sufficiently wide.  The simulations offer detailed
information about several experimentally difficult issues, in particular
the role of hydrodynamic correlations in speeding-up the
translocation process.

A three-dimensional box of size $N_x h \times N_y h \times N_z h$ 
lattice units, with $h = \Delta x$ the spacing between lattice points,
contains the solvent and the polymer.
We take $N_x = 2 N_y$, $N_y = N_z$; a separating wall is located in 
the mid-section of the $x$ direction, $x=h N_x/2$.  
We use $N_x = 100$ and $N_0 = 400$, where $N_0$
is the total number of beads in the polymer.
At the center of the separating wall, a cylindrical
hole of length $l_{hole}=10 h$ and diameter $d_{p}$ is opened.
Three different pore sizes ($d_p=5h,9h,17h$) have been used in 
the current simulations.
Translocation is induced by a constant electric field 
acting along the $x$ direction and confined to a cylindrical
channel of the same size as the hole, and length $l_p = 12h$ along the 
streamwise ($x$) direction. 
All parameters are measured in units of the lattice Boltzmann
time step and spacing, $ \Delta t$ and  $\Delta x$, respectively,
which are both set equal to $1$.
The MD time step is five times smaller than $ \Delta t$.
The pulling force associated with the electric field
in the experiments is $q_eE=0.02$ and the temperature is $k_B T/m=10^{-4}$.
The monomers interact through a Lennard-Jones 6-12 potential with parameters
$\sigma=1.8 $, and $\epsilon= \times 10^{-4}$ and
the bond length among the beads is set at $b=1.2$. 
The solvent is set at a density $\rho_{LB}=1$, with
a kinematic viscosity $\nu_{LB}=0.1$ and a drag coefficient $\gamma=0.1$.

We chose the separation $d$ between the beads to be equal to the persistence length of
double-stranded DNA, that is 50 nm, and define the lattice spacing to be $d/1.2 =40$ nm.
The hole diameters is 3 $\Delta x$. The repulsive interaction between the beads and the wall 
(with parameter $\sigma_w=1.5~\Delta x$ \cite{WCA}) leaves an effective hole of size equal to $\sim 5$ nm.
Having set the value of $\Delta x$, we choose the time step so that the kinematic viscosity is expressed as:
$\nu_w = \nu_{LB} \frac{\Delta x^2}{\Delta t}$, with $\nu_w$ the viscosity of water
($ 10^{-6} \; m^2/s$) and $\nu_{LB}$ the numerical value of the viscosity in LB units;
this procedure gives  $\Delta t \sim 160 \; ps$, with  $\nu_{LB}=0.1$.
In order to ensure numerical stability, the relation $\gamma \Delta t < 1$ must be satisfied.
Having established the value of $\Delta t$, we need to adjust the value of
the drag coefficient accordingly,  $\gamma  < 6\cdot 10^{9} sec^{-1}$. This is 
significantly smaller than an estimate of the friction based on Stoke's law for DNA \cite{LU}, 
which is equivalent to an underdamped system, or an artificially inflated bead mass.
This approach is consistent with the coarse graining of the time evolution in the coupled LB-MD scheme.

We focus on the {\it fast} translocation regime, in
which the translocation time, $t_x$, is much smaller than the
Zimm time, which is the typical relaxation time of the 
polymer towards its native (minimum energy, maximum entropy) configuration. 
This corresponds to the strong-field condition $q_e E b/kT >1$.
In this regime, simple one-dimensional Brownian models 
\cite{KARDAR}, or Fokker-Planck representations, cannot
apply because the various monomer units do not have
time to de-correlate before completing translocation.
The ensemble of simulations is generated by different realizations
of the initial polymer configuration, to account for the 
statistical nature of the process. 
Initially, the polymer
is generated by a three-dimensional random walk algorithm with
different random numbers for each polymer configuration and one bead chosen 
randomly constrained at the pore entrance. Then, the
polymer is allowed to relax for $\sim 10^4$ molecular dynamics
steps without including the fluid solvent in the relaxation, while
keeping the bead at the pore entrance fixed.
We define as time zero ($t=0$), the time after the relaxation, when the
fluid motion is also added, the pulling force begins to act and the
translocation process is initiated; at this moment the bead at the pore entrance is
also allowed to move.
At this stage we do not include any electrostatic interactions
within our model for reasons of computational simplicity.
As far as the biopolymer motion in the bulk of the solvent is concerned,
this may actually be a good approximation of experimental conditions with high salt 
concentration, which leads to strong screening of electrostatic interactions.
The situation at the pore region may require more refined treatment, beyond the scope of
the present work.

\begin{figure}
\begin{center}
\includegraphics[width=0.5\textwidth]{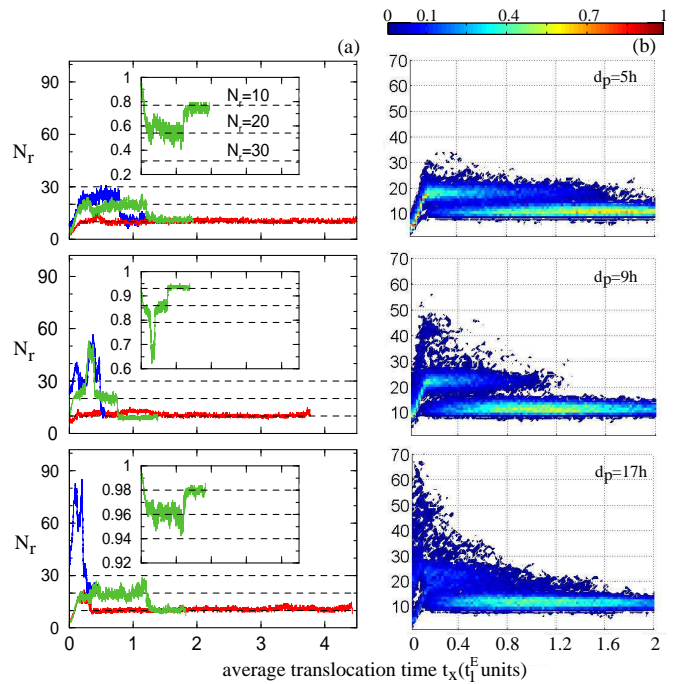}
\caption{Number of resident beads with time for three different
pore sizes $d_p=5h, 9h, 17h$ ($h=$ lattice spacing) and $N_0=400$.
(a) The fastest (minimum time, blue),
slowest (maximum time, red) and average speed (most probable time, green)
translocation events;
the insets show the current blockade for the duration of 
an event with average speed (green) with 
the current normalized to the open pore value (1).
(b) Histogram $P(N_r,t)$ of the distribution of $N_r$ with time:
short-time trajectories show
multi-file character, reaching up to $q\sim 2-8$ in the initial stage 
of the translocation depending on the pore size; long-time trajectories show little 
departure from the single-file configuration. 
}
\label{fig:Nr_dp}
\end{center}
\end{figure}
In Fig. \ref{fig:Nr_dp}(a) we present
the number of pore-resident beads $N_r(t)$ as a function of time,
for a narrow ($d_p=5h$), mid-sized ($d_p=9h$) and large ($d_p=17h$) pore, 
with $h$ the mesh spacing of the lattice Boltzmann simulation,
for representative (fastest, slowest and average speed) trajectories.
Simulations are repeated over an ensemble of $400$ realizations of different
initial conditions and for total polymer lengths up to $N_0=400$.
Time is measured in units of $t_{1}^{E}$, the time it would take for the polymer to 
translocate if the monomers were to proceed in single-file configuration 
at the drift speed; this speed is given by $v_E=q_e E /\gamma m$, with $q_e$ and 
$m$ the charge and mass of the monomer, $E$ the external electric field and
$\gamma$ the hydrodynamic drag. 
This gives $t_{1}^{E} = bN_0/v_E=12 N_0$ and the number of monomers in the pore for 
single file translocation is $N_1=10$ for the parameters used here.

Fig. \ref{fig:Nr_dp}(a) clearly shows the highly non-linear dynamics of the translocation process:
In the initial stage of the translocation, the nanopore
gets populated, with the number of resident monomers significantly overshooting 
the single-file value $N_1$, 
the horizontal dashed lines at heights $qN_1$  indicating $q$-file ($q=1,2,3,\dots$) translocation. 
The range of $q$ explored by the translocation
trajectories grows approximately with the cross-section of the pore,
going from $q \sim 2$ for the smallest pore $d_p=5h$ up to $q \sim 8$ for the 
largest one $d_p=17h$.  
Note that these values correspond to about half the maximum allowed
q-number, $q_{max} \sim d_p/b$.
The fastest events correspond to the largest $q$ value observed,
while the slowest events correspond to essentially $q=1$ throughout the translocation.
It is also noteworthy
that the translocation time typically exceeds the single-file value, $t_{1}^{E}$, 
except for the fastest events; for the most probable events $q \sim 2$ for all pore sizes
 indicating that conservative monomer-monomer interactions produce an effective slow
down compared to a single Langevin particle
subject to a constant electric drive and frictional drag $\gamma$.

Fig. \ref{fig:Nr_dp}(a) also presents the current blockade
in all three pores for the most probable event in each case,
which is the event with a translocation time close to the peak of the 
distribution over all translocation times. The current blockade
is proportional to the number of monomers in the pore
per unit area and appears to occur in well defined steps (quantized).
Specifically, these blockades are calculated from the difference
between the area of the resident beads, $\pi(\sigma/2)^2$, and the total area of the pore,
$\pi(d_p/2)^2$.
In order to investigate the quantization of the current blockade 
we monitored the distribution of $N_r(t)$ at various time frame intervals of 100 steps. 
The resident monomers block
the current across the channel, so that $N_r(t)$ conveys a direct measure of the
current drop associated with the biopolymer passage through the nanopore.
The corresponding histograms $P(N_r,t)$ for three pore sizes
are shown in Fig. \ref{fig:Nr_dp}(b).
At early times, these histograms exhibit a multi-peaked structure, which is
a clear signature of multi-file translocation. As time passes, the 
multiple peaks
recede in favor of a single-peak distribution, close to the single-file
value $N_1=10$. This was found to be a stable-attractor for every 
simulated trajectory, indicating that the tail of the polymer always
translocates as a single-file.

\begin{figure}
\begin{center}
\includegraphics[width=0.45\textwidth]{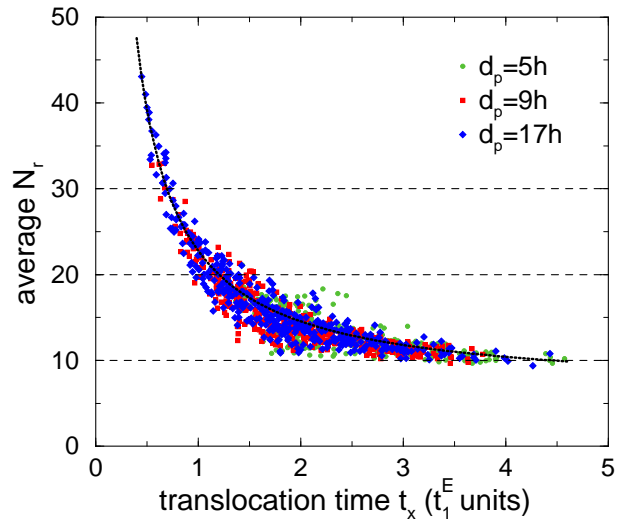}
\caption{Scatterplot of the average resident number $N_r$ versus 
translocation time (in units of $t_{1}^{E}$)  for the ensemble of translocation events 
for three values of the pore diameter, $d_p=5h,9h,17h$ and $N_0=400$.
}
\label{fig:averRr}
\end{center}
\end{figure}
Collecting all results for the average number of resident monomers $\bar{N_r}$ 
as a function of the translocation time $t_x$, for the three pore sizes studied, we find 
a simple relationship, shown in Fig. \ref{fig:averRr}.  
Experiments \cite{Li_NatMat2003} have reported that the average 
number of resident atoms $\bar{N}_r$ in each 
translocation event varies approximately inversely with the duration of translocation $t_x$:
\begin{equation}
\label{NTC}
\int_{0}^{t_x} N_r(t) \; dt \equiv  \bar {N_r} t_x  \propto N_0 = \rm{const.}
\end{equation}
The single-file asymptote $N_1=10$ ($q=1$) at long-times, $t\gg t_{1}^{E}$, is evident.
The short-time asymptote, reaching up to $4<q<5$, corresponds 
to ultrafast translocations ($t<t_{1}^{E}$) 
occurring in the case of the large-diameter pore, $d_p=17h$.
These results are intuitively reasonable, since large resident numbers imply
that more monomers cross the pore per unit time, hence the translocation
becomes faster. The results also support the notion of 
$N_r(t)$ as a measure of the time-rate of the translocation, $dN_T/dt \propto N_r$, 
from which the inverse-proportionality
between $\bar N_r$ and $t_x$ is a direct consequence of
$\int_0^{t_x} [dN_T/dt] \; dt = N_0 = \rm{const.}$
In this expression, $N_T(t)$ is the number of translocated monomers at time $t$.

\begin{figure}
\includegraphics[width=0.45\textwidth]{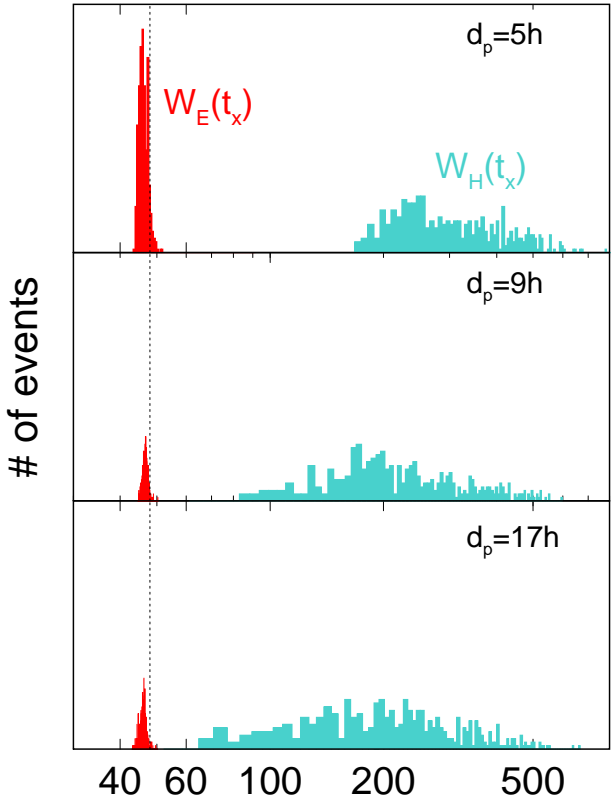}
\caption{Statistical distribution of the work performed
by the hydrodynamic and electric field during translocation events.
The vertical dotted line corresponds to the work $W_{1}^{E}$ done by the electric
field on polymers that translocate single-file.\label{pdfwork}}
\end{figure}

The simulations reveal that solvent {\it correlated} 
motion makes a substantial contribution to the translocation energetics. 
The role of hydrodynamic correlations is best highlighted by computing the work
done by the moving fluid on the polymer (we call this the {\it synergy},
$W_{H}$) over the entire translocation 
process as compared to the case of a passive fluid at rest:
\begin{equation}
W_{H}(t_x) = \gamma \int_0^{t_x} \sum_{i=1}^{N_0} \vec{u}_i(t) \cdot \vec{v}_i(t) \; dt
\label{eq:W_Hdef}
\end{equation}
where $\vec{v}_i$ is the velocity of monomer $i$ and $\vec{u}_i$ is 
the fluid velocity at the position of monomer $i$.
For the sake of comparison, it is also instructive to contrast $W_H$ with 
the corresponding work done by the electric field 
\begin{equation}
W_{E}(t_x) = q_e  \int_0^{t_x} \sum_{i=1}^{N_r(t)} \vec{E} \cdot \vec{v}_{i}(t) \; dt
\label{eq:W_Edef}
\end{equation}
where the sum extends over the resident monomers only, since the
electric field is applied at the pore region only.
These statistically averaged 
values of $W_H$ and $W_E$ reveal a number of interesting features
(see Fig. \ref{pdfwork}).
First, $W_H$ is always positive, clearly showing that hydrodynamic 
correlations provide a {\it cooperative background}, as compared to the 
case of a passive ``ether'' medium ($\vec{u}=0$).
Second, we observe that the $W_E$ has a much narrower
distribution of values than $W_H$, reflecting the ordered structure of the
biopolymer as it passes through the nanopore, as compared to its 
off-pore morphology.
It is useful to introduce the work done by the electric field on 
molecules which translocate single-file and proceed through the pore at speed 
$v_E$, $W_{1}^{E} = q_e E N_1 v_E t_{1}^{E} = b q_eE N_1 N_0$.
In the absence of any other interaction, a $q$-file 
translocation at speed
$v_E$ would complete in a time $t_x (q)=t_{1}^{E}/q$ under an electric work $q W_{1}^{E}$.
In the present simulations, $W_{1}^{E} = 0.12 N_0$, thereby $W_{1}^{E}=48$ for $N_0=400$.
Interestingly, the distribution of $W_E$ values is highly peaked 
at a value very close to $W_{1}^{E}$.
The observation that $W^E\sim W_1^E$ implies that $qv_x(q) t_x(q) \simeq v_E t_1^E$
and since the simulations show that $t_x(q) > t_1^E$, the conclusion is that  $v_x(q) < v_E /q$, 
indicating that collective motion of the monomers slows down the process.

\begin{figure}
\begin{center}
\includegraphics[width=0.47\textwidth]{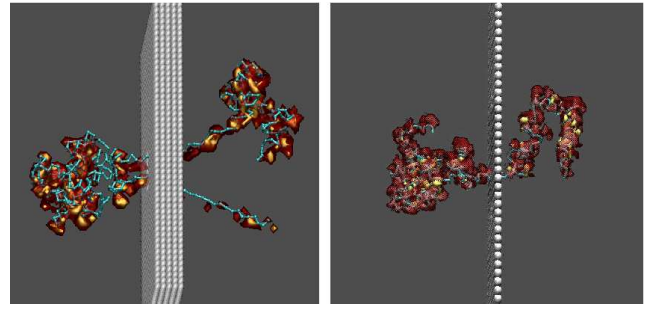}
\caption{Left panel: a typical two-folded polymer configuration ($d_p=9h$), 
at time where 65\% of the $N_0=400$ total beads 
have already translocated from right to left; colored contours show
the magnitude of the corresponding hydrodynamic synergy field
(only five of the nine wall-layers are shown).
Right panel: a single-file translocation 
event for a narrow and shallow pore ($d_p=3h, l_p=1h$) with 60\% of the
beads translocated and the corresponding magnitude of the synergy.
}
\label{fig:2fold}
\end{center}
\end{figure}
A major asset of numerical simulations for the
study of translocation processes is the direct access to visualization of 
the morphology of the translocating chain.
As an example, we show in Fig. \ref{fig:2fold}
a typical ``snapshot'' at a time when about
$65\%$ of the monomers have already passed through the pore 
of a translocating  2-folded chain of $N_0=400$ beads. 
In the same figure we show for comparison an event for the same length,
but for single-file translocation 
(unfolded chain) through a very narrow 
($d_p=3h$) and shallow ($l_p=1$) pore.
In addition to the polymer 
conformation, we show isocontours of the magnitude of the hydrodynamic synergy 
density
\begin{equation}
w_H(\vec{r};t) = \gamma \sum_{i\in B(\vec{r})} \vec{u}_i(t) \cdot \vec{v}_i(t)
\end{equation} 
which is a local (in both space and time) version of the total synergy $W_H$ defined
in Eq.(\ref{eq:W_Hdef}), with 
$B(\vec{r})$ a grid cell centered around
location $\vec{r}=(x,y,z)$.
The contours of $w_H(\vec{r})$ illustrate the cooperative nature of 
the hydrodynamic field, with regions of high co-moving flow 
surrounding the translocating polymer and assisting its motion.
This is suggestive of the notion of an ``effective'' polymer, dressed
with the hydrodynamic synergy field, which acts as a self-consistent
lubricant, helping the polymer to negotiate a faster passage through
the nanopore.

\begin{figure}
\begin{center}
\includegraphics[width=0.45\textwidth]{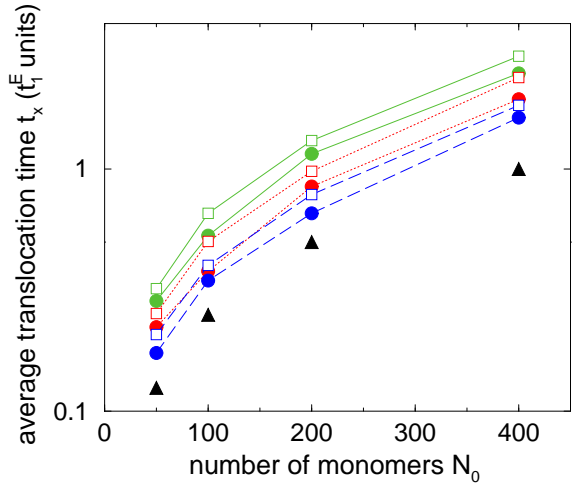}
\caption{Average translocation time as a function of polymer length,
with (closed circles) and without (open squares) hydrodynamic 
interactions. Colors correspond to different pore diameters $d_p= 5h$ 
(green solid line), $9h$ (red dotted line), and $17h$ (blue dashed line).
Black triangles indicate the value of the single-file translocation time $t_{1}^{E}$
for each value of $N_0$ with hydrodynamics 
(all numbers are scaled to the value of $t_{1}^{E}$ for $N_0=400$).
}
\label{fig:timeNpore}
\end{center}
\end{figure}
We further investigate this issue by inspecting the
average (over an ensemble of $400$ realizations) translocation time, 
$\langle t_x \rangle$, as a function of the polymer length, with and without 
hydrodynamics. The results are shown in Fig. \ref{fig:timeNpore}: hydrodynamics consistently
accelerates the translocation by roughly $30\%$ percent. 
More intuitively, hydrodynamics literally re-normalizes the diameter of the
pore: as is clearly visible in Fig. \ref{fig:timeNpore}, a pore of diameter
$d_{p}=5h$ for a bare polymer (without the hydrodynamic field) 
is essentially equivalent to a pore of almost
double diameter $d_{p}=9h$ for the hydrodynamically-dressed polymer.
In order to assess the degree of correlation between the translocation
dynamics of the ``dressed'' polymer versus the actual one, we
have measured the translocated specific synergy (synergy per monomer), defined as
\begin{equation}
\langle w_H (t) \rangle = \frac{\gamma}{N_T(t)} \sum_{i=1}^{N_T(t)} 
\vec{v}_i(t) \cdot \vec{u}_i(t)
\end{equation} 	 
with $N_T(t)$ the number of translocated monomers at time $t$.
Clearly, any implicit time-dependent functional dependence of the form 
$\langle w_H (t)\rangle = \langle w_H (N_T(t)) \rangle$ 
would indicate that mass and synergy translocate in a synchronized manner.
We find that the ratio $\langle\vec{u}\cdot\vec{\upsilon}\rangle/kT \sim 5$, 
reflecting the fact that the solvent locally ``follows'' the monomer and providing
a measure of the relative importance of synergistic versus thermal forces.
Our results show that $\langle w_H(t)\rangle$ is essentially
constant throughout the translocation process. 
This implies a direct proportionality between the translocated synergy 
and the number of translocated beads and  supports the notion that the
``dressed'' and the actual polymer proceed in full 
synchronization across the nanopore. 

In conclusion, by using a new multiscale methodology based on the
direct coupling of constrained molecular dynamics for the solute
biopolymers with a lattice Boltzmann treatment of solvent dynamics, we
have been able to confirm a number of experimental
observations, such as a direct relation between quantized current
blockades and multi-folded polymer conformations during the
translocation process. 
In particular, the simulations reveal an intimate
connection between polymer and hydrodynamic motion which 
promotes a cooperative background for the translocating molecule, 
thus resulting in 
a significant acceleration of the translocation process. 
Such an acceleration can also be interpreted as the outcome of a 
renormalization 
of the actual polymer geometry into an effective one, more conducive 
to translocation.
This opens up exciting prospects for the development of optimized 
nano-hydrodynamic devices based
on the fine-tuning of hydrodynamic correlations.
As an example, one may envisage multi-translocation chips, whereby 
multiple molecules
would translocate in parallel across membranes with an array of pores.
The optimization of such devices will require control of solvent-mediated
molecule-molecule interactions to minimize destructive interference 
between translocation events.

{\bf Acknowledgements}

MB, SM and SS acknowledge support from the Initiative
for Innovative Computing and thank 
the Physics Department of Harvard University for its hospitality. 
MF acknowledges support by the Nanoscale Science
and Engineering Center of the National Science Foundation
under NSF Award Number PHY-0117795.


\end{document}